\documentstyle[epsfig]{l-aa}

\begin{document}

\def\ltsima{$\; \buildrel < \over \sim \;$}
\def\simlt{\lower.5ex\hbox{\ltsima}}

%\large
   \thesaurus{ 4(11.09.1: Mrk 509, MCG+8-11-11;  % Galaxies: individual: 
% Mrk 509, MCG+8-11-11,
                11.19.1; % Galaxies: Seyfert,
                13.25.2)} % X-rays: galaxies

   \title{BeppoSAX observations of Mrk~509 and MCG~+8-11-11}

   \author{G. C. Perola
          \inst{1}
    \and G. Matt\
          \inst{1}
\and F. Fiore
          \inst{2}
\and P. Grandi
          \inst{3}
\and M. Guainazzi
          \inst{4}
\and F. Haardt
          \inst{5}
\and L. Maraschi
          \inst{6}
\and T. Mineo
          \inst{7}
\and F. Nicastro
          \inst{8}
\and L. Piro
          \inst{3}
          }
   \offprints{G.C. Perola, perola@fis.uniroma3.it}
% \\
%              proposed choice: Main Journal \\
%              proposed section: Extragalactic astronomy \\
%              proofs to: G.C. Perola, tel.~+39--6--55177022
%fax~+39--6--5579303,
%                         email: matt@fis.uniroma3.it}

   \institute{Dipartimento di Fisica, Universit\`a degli Studi ``Roma Tre",
              Via della Vasca Navale 84, I--00146 Roma, Italy
   \and Osservatorio Astronomico di Roma, Via dell'Osservatorio,
        I--00044 Monteporzio Catone, Italy
  \and Istituto di Astrofisica Spaziale, C.N.R., Via Fosso del Cavaliere,
                I--00133 Roma, Italy
   \and XMM SOC, VILSPA--ESA, Apartado 50727, E--28080 Madrid, Spain
   \and Dipartimento di Scienze, Universit\`a dell'Insubria, Via Lucini 3,
         I--22100 Como, Italy
   \and Osservatorio Astronomico di Brera, Via Brera 28, I--20121 Milano, Italy
   \and Istituto di Fisica Cosmica ed Applicazioni all'Informatica, C.N.R.,
        Via U. La Malfa 153, I--90146 Palermo, Italy
     \and Harvard-Smithsonian Center for Astrophysics,
                60 Garden st., Cambridge MA 02138 USA
     }

   \date{Received February 8, 2000 / Accepted }

   \maketitle

\markboth{G.C. Perola et al.: BeppoSAX observations of Mrk~509 
and MCG~+8-11-11}{}

    \begin{abstract}
BeppoSAX observations of the Seyfert galaxies Mrk 509 and
MCG~+8-11-11 are presented. Earlier evidence of a soft
excess in Mrk 509 is confirmed. This excess is found to be
better represented by a power law than by a black body:
with a photon slope $\Gamma_s$ of 2.5, its extrapolation 
matches the flux recorded in the far UV. An ASCA observation,
which appeared to exclude the presence of the excess while
showing instead evidence of a warm absorber, turns out to
be compatible with the coexistence of the excess seen with
BeppoSAX and of the warm absorber. The hard power law of
Mrk 509 is seen for the first time to be affected by a
cut--off at high energies, with an e--folding energy 
of about 70 keV. In MCG~+8-11-11 the cut--off is found
at about 170 keV, consistent within the combined errors with a
previous estimate from a ASCA+OSSE/CGRO observation.
In both objects the reflection component is clearly detected.
In Mrk 509 its strength, together with that of the iron K line,
indicates a solid angle $\Omega$, subtended by the reprocessing
gas in the accretion disk, much less than 2$\pi$, but lacking
a valid constraint on the inclination angle this evidence
is not as convincing as that found with BeppoSAX in IC 4329A.
In MCG~+8-11-11 the same parameters
are instead consistent with $\Omega$=2$\pi$, but the comparison with
an ASCA observation, when the flux level was about
2.5 times weaker, suggests that a substantial fraction of the
angle might be associated with gas farther out than the accretion disk.  

   \end{abstract}

\section{Introduction}

As part of a broad band spectral survey, with the BeppoSAX 
satellite, of a sample of bright Seyfert 1 galaxies (2--10 keV flux
greater than 1--2 mCrab), this paper follows those on
NGC 4593 (Guainazzi et al. 1999) and IC 4329A (Perola
et al. 1999), where the goals of the program are described
in detail. In short terms, the survey aims at determining
the parameters of the various spectral components typical
of Seyfert 1s, and especially of the broad ones, with better
confidence than could be done before, by exploiting the
opportunity to cover simultaneously the band 0.1--200 keV,
offered by the Narrow Field Instruments
(NFI) onboard BeppoSAX.

This paper reports on the observations of two objects,
namely Mrk 509 and MCG +8-11-11. Mrk 509 is a Seyfert 1.0
galaxy at $z$=0.0344 with a luminosity
$L$(2--10 keV) $\sim 3\times10^{44}$ erg s$^{-1}$ ($H_o$= 50 km s$^{-1}$ 
Mpc$^{-1}$). The galactic interstellar column toward
this object amounts to $N_{H,g}$=4.4$\times10^{20}$ cm$^{-2}$ 
(Murphy et al. 1996).

MCG +8-11-11 is a Seyfert 1.5 galaxy at $z$=0.0205 with a
luminosity $L$(2-10 keV)$\sim10^{44}$ erg s$^{-1}$.
Because of the low galactic latitude (b=10.4$^{\circ}$) its spectrum
is absorbed by a large galactic column, $N_{H,g}$=2.03$\times10^{21}$ cm$^{-2}$
(Elvis et al. 1989).

Special reasons of interest from earlier observations
of the two objects are the following. In Mrk 509 the
report of a sizeable soft excess below about 1 keV
in Ginga and ROSAT simultaneous observations by
Pounds et al. (1994), which appeared to confirm previous
claims in data from HEAO-1 (Singh et al. 1985), EXOSAT
(Morini et al. 1987) and Ginga (Singh et al. 1990),
was itself not confirmed in one ASCA observation
by Reynolds (1997) and by George et al. (1998), 
who claim instead evidence of a warm
absorber. In MCG +8-11-11 a high energy cut--off in the power law
was detected in simultaneous ASCA and OSSE/CGRO
observations at about 270 keV (Grandi et al. 1998).

Observations and data reduction are described in Sect. 2,
the spectral analysis in Sect. 3. The results are commented
and compared with previous ones in Sect. 4, and
the conclusions are drawn in Sect. 5.

\section{Observations and data reduction}

The data used in this paper were collected from three
of the four NFI on the Italian-Dutch satellite BeppoSAX
(Boella et al., 1997a), namely the imaging instruments LECS 
(0.1--10 keV, Parmar et al. 1997) and MECS (1.8--10.5 keV,
Boella et al. 1997b), and the collimated instrument
PDS (13--200 keV, Frontera et al. 1997),
with the collimator in the rocking mode to monitor the background.
Because of remaining uncertainties in the LECS
data above 4 keV, these will be ignored in the spectral analysis.

\begin{table*}
\centering
\caption{Observation epochs and mean count rates. The count rate for the MECS
refers to two units}
\vspace{0.05in}
\begin{tabular}{|c|c|cc|cc|cc|}
\hline
~& ~ & ~ & ~ & ~ & ~ & ~ & ~\cr
~ & Start date  & LECS  & LECS & MECS & MECS  & PDS & PDS\cr
\hline
~ & ~ & t$_{\rm exp}$ (s) & CR (cts~s$^{-1}$) & t$_{\rm exp}$ (s) & CR (cts~s$^{-1}$) 
& t$_{\rm exp}$ (s) & CR (cts~s$^{-1}$)  \cr
~& ~ & ~ & (0.1-4~keV) & ~ & (1.8-10.5~keV) & ~ & (13-200~keV) \cr
\noalign {\hrule}
\hline
~& ~ & ~ & ~ & ~ & ~ & ~ & ~\cr
Mrk~509~(part~1) & 1998-May-18 & 23598 & 0.464$\pm$0.004 
& 51864 & 0.601$\pm$0.003 & 48102 & 0.73$\pm$0.04 \cr
~& (12h~22m~24s~UT) & ~ & ~ & ~ & ~ & ~ & ~\cr
~& ~ & ~ & ~ & ~ & ~ & ~ & ~\cr
Mrk~509~(part~2) & 1998-Oct-11 & 18188 & 0.590$\pm$0.006 
& 35970 & 0.736$\pm$0.005 & 34000 & 0.87$\pm$0.04 \cr
~& (04h~34m~07s~UT) & ~ & ~ & ~ & ~ & ~ & ~\cr
~& ~ & ~ & ~ & ~ & ~ & ~ & ~\cr
MCG~+8-11-11& 1998-Oct-7 & 29840 & 0.346$\pm$0.003 
& 74866 & 0.637$\pm$0.003 & 72472 & 0.98$\pm$0.04 \cr
~& (23h~10m~45s~UT) & ~ & ~ & ~ & ~ & ~ & ~\cr
~& ~ & ~ & ~ & ~ & ~ & ~ & ~\cr
\hline
\end{tabular}
\end{table*}

In Table 1 start dates of the observations, net exposure
times and net average count rates are given. While those
of MECS and PDS are comparable, the exposure times of the LECS 
are much shorter because this instrument is operated only
during the night time fraction of the orbit. The reduction
procedures and screening criteria adopted are standard
(see Guainazzi et al. 1999). In particular, of the two options
available for the PDS data, the Rise Time selection was used.

The spectral counts  were extracted from 
a circular region of 4' and 8' radius around the source centroid
in the MECS and LECS images respectively. The background
was subtracted using spectra from blank sky files in the same
position of the detectors. Due to the low galactic latitude
of MCG +8-11-11, this procedure overestimates the background
below about 0.4 keV: ignoring the counts below 0.4 keV, as we
shall do, is of no practical consequence in the spectral fitting,
due to the large value of $N_{H,g}$.

Spectra and light curves from the PDS were obtained from direct
subtraction of the off- from the on-source products. The enquiry
on the presence of sources that could significantly
contaminate the target signal in the 1.3$^{\circ}$ FWHM field
of view gave negative results. 

During the observations both objects underwent intensity
variations, whose detailed study goes beyond the scope of this paper.
A hardness ratio analysis of the counts shows at best marginal
evidence of spectral variations, hence we feel justified 
in applying the spectral analysis to the integrated
counts. In particular the two parts of the observation
of Mrk 509 will be combined. To take care of the inhomogeneity
in the time coverage between LECS and MECS, we shall
let their relative normalization as a free parameter.
The normalization of the PDS to the MECS will instead be held fixed.

\section{Spectral analysis}

The analysis of the integral spectral counts, from
the three instruments together, was performed with 
XSPEC (version 10). In the first step a ``Baseline Model
Spectrum" (BMS) is adopted, composed of: a Power Law with
an exponential cut--off (PL), $A E^{-\Gamma} \exp(-E/E_f)$;
a Reflection Component (RC), with $r=\Omega/2\pi$, the
solid angle fraction of a neutral, plane parallel slab
illuminated by the PL photons, and with the inclination
angle $i$ set, for reference, equal to 0$^{\circ}$  (this
angle, if left free, is too loosely constrained, see
Perola et al. 1999); a uniform neutral column of absorbing gas 
at the object redshift, N$_H$, in addition to the galactic
$N_{H,g}$ (in both slab and column
the element abundances are the cosmic values in
Anders \& Grevesse 1993); a gaussian iron K line, with 
energy $E_k$, width $\sigma_k$ and intensity $I_k$ (also given as
an EW). Further steps include, in both objects: 
the addition of the O{\sc vii}
(0.74 keV) and O{\sc viii} (0.87 keV) edges, with their optical
depths $\tau$ as free parameters, as the most direct way to verify 
the existence of a warm absorber (WA); the relativistic
description of the iron line profile; in Mrk 509 the addition of 
a soft broad component.

The energy bins chosen represent about one third of
the instrumental resolution, which is a function of the
energy. The normalization C of the PDS relative to the MECS
was adopted equal to 0.8 (Fiore et al. 1999). The ``statistical"
errors correspond to
the 90\% confidence interval for two interesting parameters,
or $\Delta\chi^2$=4.61. The energy $E_k$ is always given as in
the frame of the host galaxy.

\begin{table*}
\centering
\caption{Spectral fits (cos~$i$=1)}
\vspace{0.05in}
\begin{tabular}{|c|c|c|c|}
\hline
~ & ~ & ~ & ~ \cr
~ & Mrk~509 & Mrk~509 & MCG~+8-11-11 \cr
~ & ~ & ~ & ~ \cr
\hline
~ & ~ & ~ & ~ \cr
~ & BMS + WA & BMS + SPL + WA & BMS + WA \cr
~ & ~ & ~ & ~ \cr
\hline
~ & ~ & ~ & ~ \cr
F(2-10~keV)$^a$ & 5.66$\pm$0.03 & 5.66$\pm$0.03 & 5.61$\pm$0.03 \cr
N$_H$~(10$^{20}$~cm$^{-2}$)$^b$ & $<$0.1 & 0.1$^{+0.4}_{-0.1}$ & $<$2.3\cr
A~(10$^{-2}$~cm$^{-2}$~s$^{-1}$~keV$^{-1}$) & 2.12 & 0.89 & 1.66 \cr
$\Gamma$ & 2.11$^{+0.02}_{-0.03}$ & 1.58$^{+0.08}_{-0.09}$$^d$ & 
1.84$\pm$0.05 \cr
E$_f$~(keV) & 302$^{+1564}_{-162}$ & 67$^{+30}_{-20}$$^d$ & 169$^{+318}_{-78}$ \cr
$r$ & 1.99$^{+0.51}_{-0.42}$ & 0.57$^{+0.34}_{-0.32}$$^d$ & 
0.98$^{+0.56}_{-0.39}$ \cr
E$_k$~(keV)$^c$ & 6.4~(fix) & 6.70$^{+0.38}_{-0.26}$$^d$ & 
6.49$^{+0.14}_{-0.13}$ \cr
$\sigma_k$~(keV) & 2.92$^{+1.20}_{-0.67}$ & 0.36$^{+0.76}_{-0.36}$$^d$ 
& 0.17$^{+0.30}_{-0.17}$ \cr
I$_k$~(10$^{-4}$~cm$^{-2}$~s$^{-1}$) & 6.1$^{+2.2}_{-1.8}$ & 
0.62$^{+0.76}_{-0.35}$$^d$ & 0.81$^{+0.45}_{-0.33}$  \cr
EW~(eV) & 980$^{+353}_{-289}$ & 107$^{+130}_{-60}$$^d$ 
& 133$^{+74}_{-54}$ \cr
$\tau$(O~{\sc vii}) & 0.15$^{+0.14}_{-0.13}$ & 0.06$^{+0.11}_{-0.06}$$^d$ &
0.17$^{+0.19}_{-0.17}$ \cr
$\tau$(O~{\sc viii}) & $<$0.17  & $<$0.08$^d$ &  $<$0.18 \cr
A$_s$~(10$^{-2}$~cm$^{-2}$~s$^{-1}$~keV$^{-1}$) & -- & 1.11 & -- \cr
$\Gamma_s$ & -- & 2.50$^{+0.94}_{-0.25}$$^e$ & -- \cr
$\chi^2$/d.o.f. & 154.9/135 & 134.5/132 & 113.7/130 \cr
~ & ~ & ~ & ~ \cr
\hline
\end{tabular}
\begin{tabular}{c}
$^a$As observed in 10$^{-11}$ erg cm$^{-2}$ s$^{-1}$. \cr
$^b$In excess of N$_{H,g}$=4.4$\times10^{20}$ cm$^{-2}$  (Mrk~509),
2.03$\times10^{21}$ cm$^{-2}$  (MCG~+8-11-11).\cr
$^c$In the frame of the host galaxy.\cr
$^d$Errors computed with N$_{H}$ and $\Gamma_s$ frozen at their best fit 
values.\cr
$^e$Errors computed with N$_{H}$ frozen at its best fit value.
\end{tabular}
\end{table*}

\subsection{Mrk 509}

Although the fit {\it without} the emission
line gives a much worse $\chi^2$, in the BMS fit
$E_k$ is ill-determined, and,
when $E_k$ is fixed at 6.4 keV (or any other value up to 6.9 keV),
the line width turns out to be very large, about
2.9 keV (with EW$\sim$950). The $\chi^2$=176.8/137, to be compared with
245.4/139 without the line.
The addition of the two O edges 
leads to a substantial improvement in $\chi^2$, but the problem with $E_k$ 
and $\sigma_k$ remains. When $E_k$ is fixed at 6.4 keV,
$\chi^2$=154.9/135. The results are given in Table 2
under BMS+WA.
Compared with the results obtained by Reynolds (1997), for the same model
except for the absence of the RC, the optical depths are consintent, 
whereas his estimate of $N_H$=2.1($\pm$0.6)$\times10^{20}$ cm$^{-2}$
is incompatible with our upper limit (the same holds with
$N_{H,g}$=4.2$\times10^{20}$ cm$^{-2}$, as adopted by Reynolds).
The other parameters remain similar 
to those of the pure BMS fit. The extremely large value
of $\sigma_k$ is likely due to a bad description of the continuum
at the line energy, which would affect also the
determination of $r$. The ratio data/model when both $I_k$ and $r$
are set to zero  (Fig. 1) indicates  that the continuum becomes
substantially harder above 2--3 keV, and  therefore
the fit manages to reproduce the data by freely adjusting $I_k$,
$\sigma_k$ and $r$. On the other hand in the same plot the
presence of a (much narrower) iron emission line
is evident. For completeness, we report that the use of the module
ABSORI in XSPEC, as a more detailed description of the warm
absorber than the two edges, leads to the same value of
$\Gamma$ (2.11), hence to the same problem with $\sigma_k$
and $r$.

\begin{figure}
\epsfig{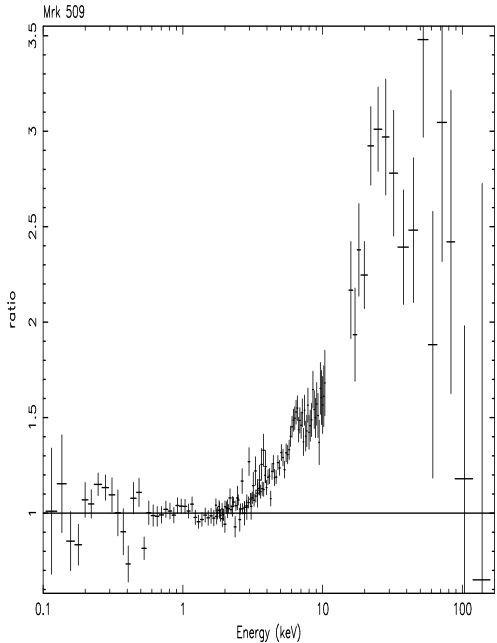}
\caption{Mrk 509. Ratio of LECS, MECS and PDS data to the model
BMS+WA, when the parameter values are those given in Table 2,
except for $I_k$ and $r$, which  are set equal to zero.}
\end{figure}

To further investigate this aspect, we fitted
the LECS data with a simple power law
absorbed by $N_{H,g}$. The $\chi^2$ is 142/53, and
the ratio data/model in Fig. 2
shows a curvature in the
slope. The curvature remains after the
addition of the two edges, which brings
the $\chi^2$ down to 115/51, a still fully unacceptable value.
Some improvement is achieved using ABSORI, but the
$\chi^2$ = 89/51 remains unacceptable.

\begin{figure}
\epsfig{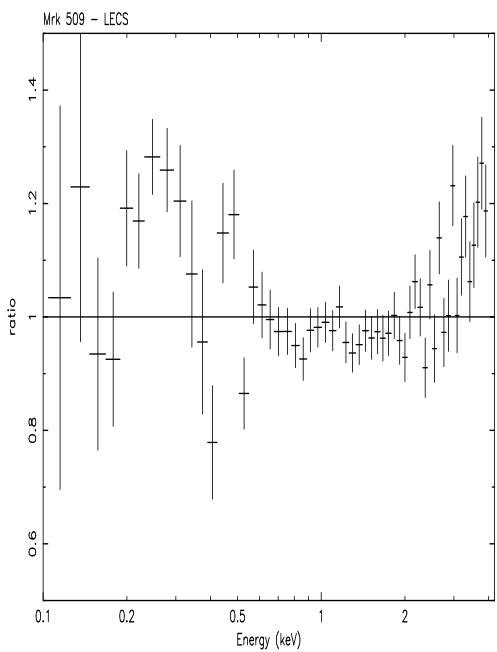}
\caption{ Mrk 509. Ratio data to model, when the LECS data
are fitted with a simple power law, absorbed by $N_{H,g}$.}
\end{figure}

We therefore decided to describe this situation by
adding to the BMS another broad component, 
referred to as a Soft Power Law (SPL, $A_s E^{-\Gamma_s}$).
The line energy is now reasonably well determined, hence this
fit includes three more free parameters relative to
the BMS. The $\chi^2$=135.6/134, and the outcome is
dramatically different: along with a much harder slope
$\Gamma$ and a much smaller reflection $r$, $\sigma_k$ is about
0.4 keV only (EW$\sim$100 eV); moreover, $E_f$ shifts from about
300 to about 70 keV. 
The addition of the two edges leads now to no
significant improvement ($\chi^2$=134.5/132) and only O{\sc vii}
is at best marginally detected.
The high significance of the line and of the 
cut--off is instead demonstrated by comparing the $\chi^2$
given above with those obtained without the line,
154.5/135, and without
both line and cut--off, 212.9/136.

The results, under BMS+SPL+WA,  are given in Table 2 and 
illustrated in Fig. 3.
Note that again $N_H$ is negligible compared to $N_{H,g}$.
To avoid their strong interplay with $\Gamma$, 
both $N_H$ and $\Gamma_s$ were frozen at their best fit values
when the errors on all the other parameters, given in 
Table 2, were determined; likewise the confidence 
contours for the couples ($\Gamma$, $r$), ($\Gamma$, $E_f$), and
($E_k$, $\sigma_k$), given in Fig. 4, 5 and 6, were also
obtained. Caution is therefore due when discussing single
parameters, in particular the marginal evidence that the
value of $E_k$ in Table 2 is larger than the neutral value
of 6.4 keV should be regarded as hardly significant. As an
independent test, we checked whether a partially ionized reflector 
was preferable, and obtained a best fit value of zero for the
ionization parameter.

\begin{figure}
\epsfig{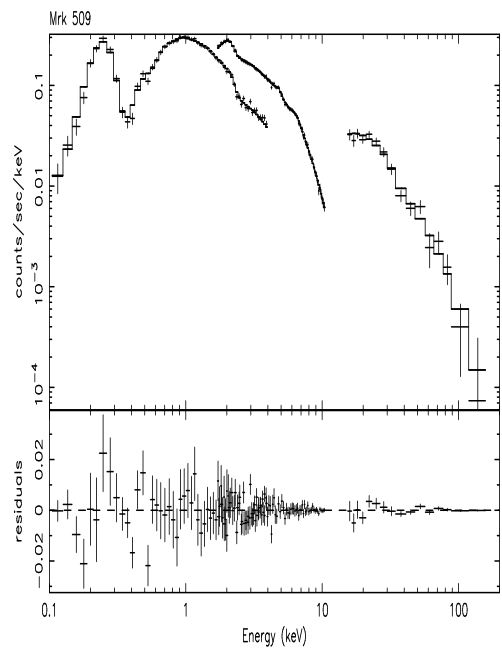}
\caption{Mrk 509. LECS, MECS and PDS spectra with the best fit
BMS+SPL+WA given in Table 2 (upper panel), and residuals (lower
panel).}
\end{figure}

\begin{figure}
\epsfig{file=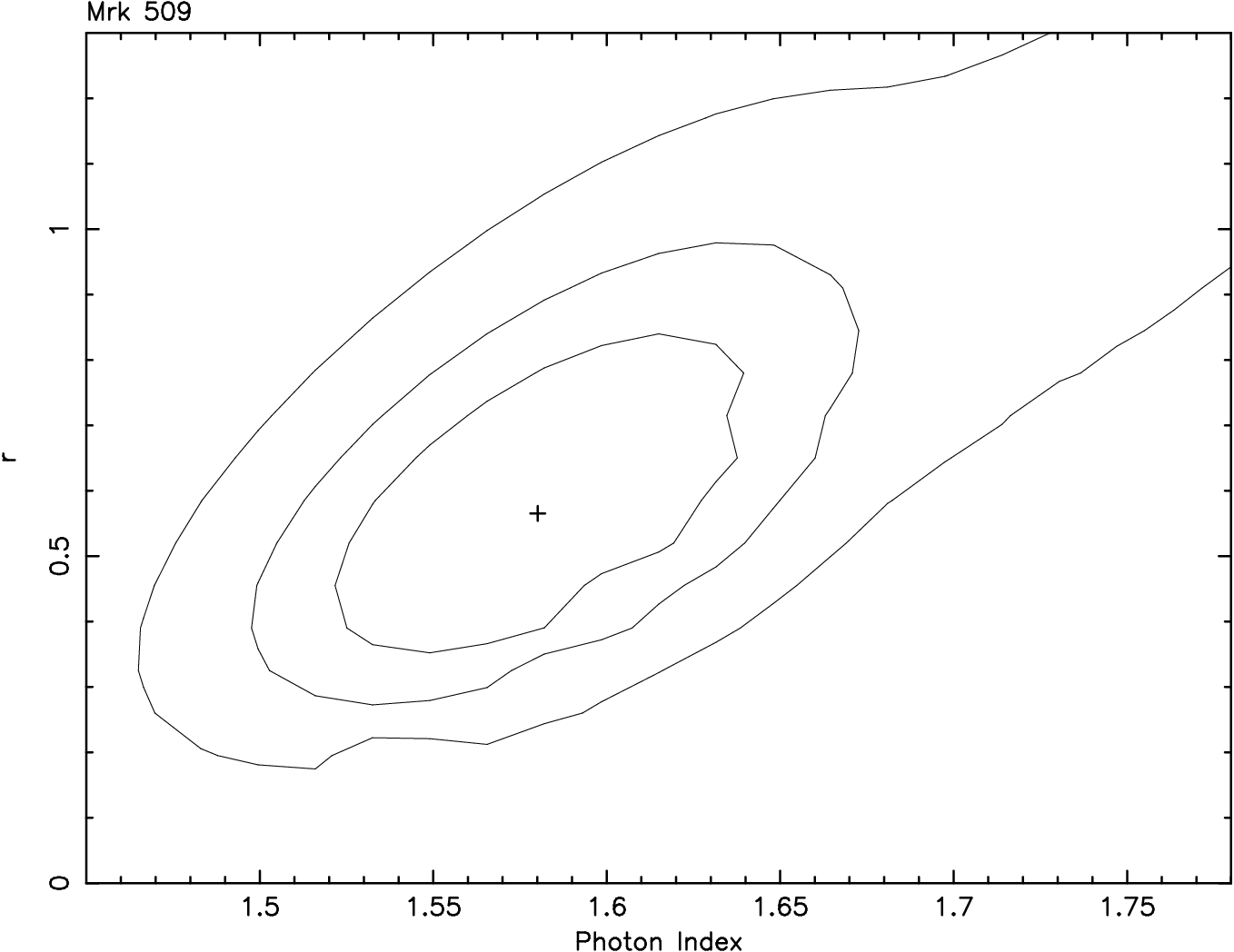, height=8.5cm,width=6.5cm, angle=-90 }
\caption{Mrk 509. Confidence (67, 90, 99\%) contours of
$\Gamma$ and $r$ in the BMS+SPL+WA fit given in Table 2,
obtained with $N_H$ and
$\Gamma_s$ frozen at their best fit values.}
\end{figure}

\begin{figure}
\epsfig{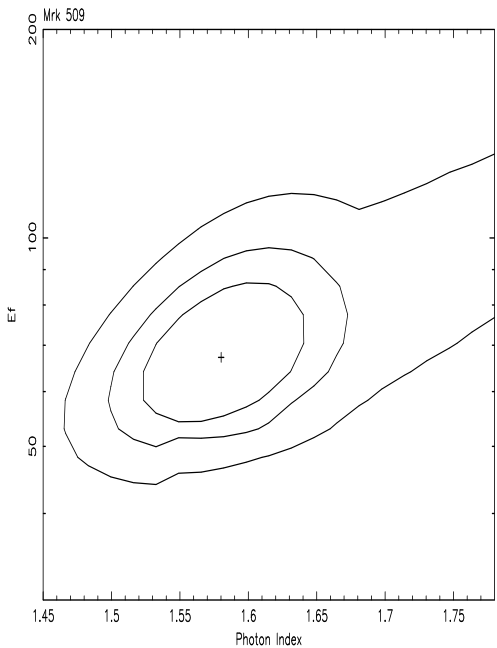}
\caption{Mrk 509. Confidence (67, 90, 99\%) contours of
$\Gamma$ and $E_f$ in the BMS+SPL+WA fit given in Table 2,
obtained with $N_H$ and
$\Gamma_s$ frozen at their best fit values. }
\end{figure}

\begin{figure}
\epsfig{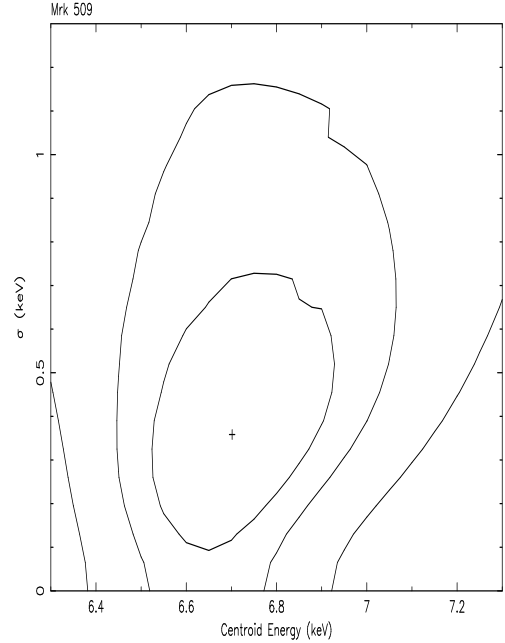}
\caption{Mrk 509. Confidence (67, 90, 99\%) contours of
$E_k$ and $\sigma_k$ in the BMS+SPL+WA fit given in Table 2,
obtained with $N_H$ and
$\Gamma_s$ frozen at their best fit values. }
\end{figure}

Since the iron line is marginally 
resolved, we tried the relativistic profile description for
an accretion disk around
a Schwarzschild black hole, with the inner radius $R_i$=6 $r_g$ and 
the emissivity slope $\beta$=-2 fixed: 
the line energy, the outer radius $R_{out}$, the intensity
and the inclination angle being the free parameters.
The angle in the RC component was linked to that of
the disk, the main goal of this exercise being
to provide a constraint on $i$. The $\chi^2$
is comparable to that of the gaussian description.
The inclination angle is 56$^{\circ}$, but at the 90\% confidence level
it could be anything between 0$^{\circ}$ and 87$^{\circ}$; this
reflects itself into a 90\% interval for $r$ ranging from 0.35 to 3.8.
In practice no useful constraint is obtained on $i$,
$R_{out}\sim$ 2900, and all other parameter values
remain very similar to those in Table 2.

The comparison between the two parts of the observation,
fitted independently, shows no significant differences in
the spectral parameters. In particular for the parameters of
the broad components in the first (second) part of the
observation we obtain: $\Gamma_s$ = 2.46$^{+2.02}_{-0.33}$
($\Gamma_s$ = 2.72$^{+1.92}_{-0.40}$), $\Gamma$ = 
1.64$^{+0.11}_{-0.17}$ ($\Gamma$ = 1.74$^{+0.22}_{-0.11}$),
$E_f$ = 79$^{+60}_{-29}$ keV ($E_f$ = 80$^{+158}_{-30}$ keV),
$r$ = 0.64$^{+0.56}_{-0.44}$ ($r$ = 0.83$^{+0.76}_{-0.47}$).
A joint fit of the two parts, with all parameters linked
(that is the same) except for the normalizations, gives
a very good $\chi^2$ = 285/277: the ratios of the soft
and of the hard PL intensities between the first and the
second part are equal to 1:1.40 and 1:1.21 respectively,
thus indicating that both components increased,
but the SPL more than the hard PL. 

As an alternative to the SPL, we tested the black body description
of the soft excess, which was adopted by Pounds et al. (1994).
The $\chi^2$ = 137.6/132 is perfectly acceptable, and $kT$ = 0.07 keV,
but, although somewhat smaller than those from the BMS+WA fit,
the values $\sigma_k$ = 2.3 keV and EW = 580 eV remain improbably large,
and lead us to believe that the SPL option is to be preferred.

\subsection{MCG +8-11-11}

The BMS fit (above 0.4 keV) yields $\chi^2$=117.5/132.
The addition of the two edges leads to a slight decrease,
$\chi^2$=113.7/130, and only OVII is marginally detected.
The results of this fit are given in Table 2 under BMS+WA
and shown in Fig. 7. They are further illustrated as confidence
contours for the couples ($\Gamma$, $r$), ($\Gamma$, $E_f$), and
($E_k$, $\sigma_k$) in Fig. 8, 9, 10.
The significance of the line and of the
cut--off is further demonstrated by the $\chi^2$
obtained without the line
(155/133) and without
both line and cut--off (189.6/134).
The iron line, consistent in energy with 6.4 keV, is
not resolved. The fit with a relativistic profile
gives essentially the same EW, and no useful information
on the inclination angle.
Albeit marginal, the detection of a warm
absorber is new in this object. Note, on the other hand, that
the intrinsic, cold $N_H$ is negligible compared to $N_{H,g}$.

\begin{figure}
\epsfig{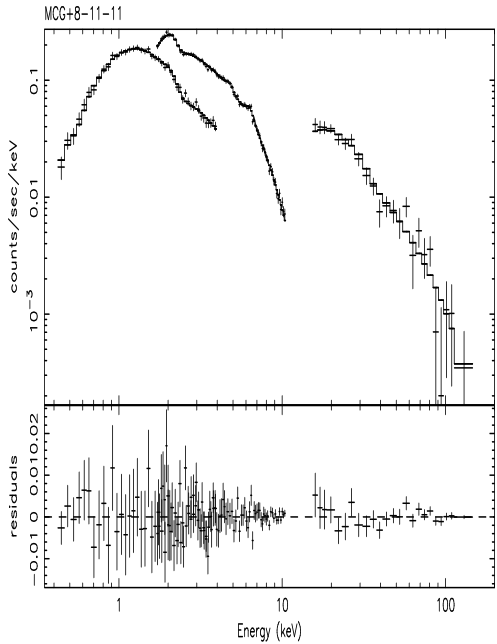}
\caption{MCG +8-11-11. LECS, MECS and PDS spectra with the best fit
BMS+WA given in Table 2 (upper panel), and residuals (lower
panel).}
\end{figure}

\begin{figure}
\epsfig{file=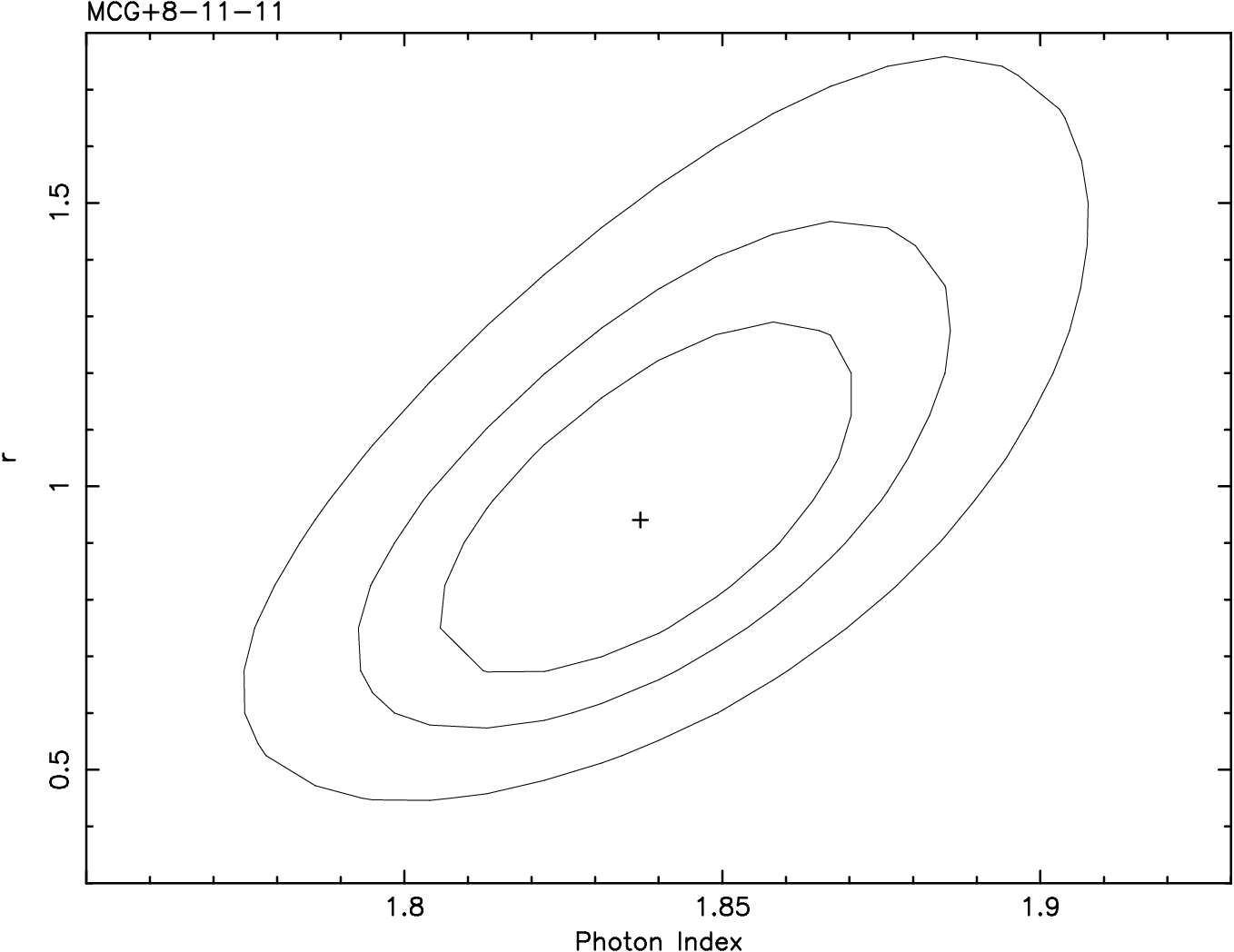, height=8.5cm,width=6.5cm, angle=-90 }
\caption{MCG +8-11-11.  Confidence (67, 90, 99\%) contours of
$\Gamma$ and $r$ in the BMS+WA fit given in Table 2.}
\end{figure}

\begin{figure}
\epsfig{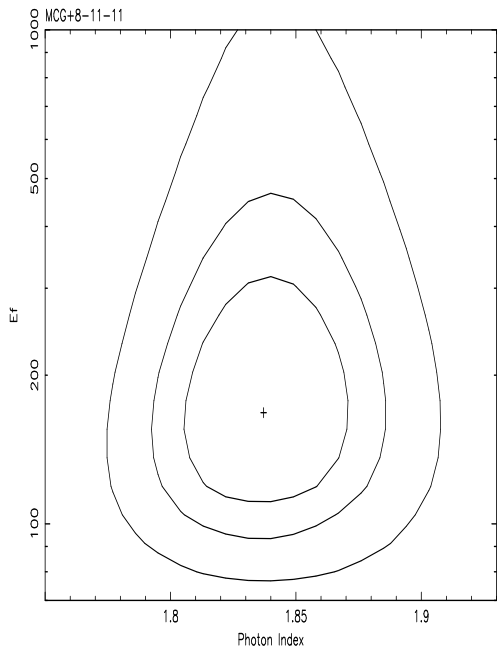}
\caption{MCG +8-11-11. Confidence (67, 90, 99\%) contours of
$\Gamma$ and $E_f$ in the BMS+WA fit given in Table 2. }
\end{figure}

\begin{figure}
\epsfig{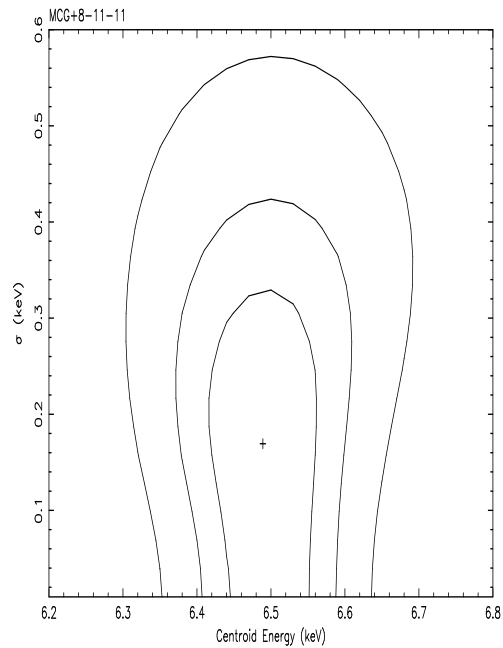}
\caption{MCG +8-11-11. Confidence (67, 90, 99\%) contours of
$E_k$ and $\sigma_k$ in the BMS+WA fit given in Table 2.}
\end{figure}

\section{Comments and comparison with previous results}

\subsection{Mrk 509}

We believe that the BeppoSAX data, thanks to their major asset,
namely the very broad band, demontrate that the continuum is
steeper at low than at high energies ($\Delta\Gamma\sim$0.9), 
thus confirming that reports, from measurements prior to
ASCA, of a ``soft excess" below about 1 keV were substantially
correct. Il would be important to establish whether the curvature is
due to the superposition of two components (as modelled by us)
or is an intrinsic property of the same physical component.
The comparison between the two parts of our observation
described in Sect. 3 suggests at least a physical connection
between the two components.
The extrapolation of the SPL to UV wavelengths is also of interest.
The reddening corrected ($E(B-V)$=0.08) $\nu f_\nu$ at 1350 A is
given in Wang et al. (1998) to be 16.2($\pm$1.5)$\times$10$^{-11}$ erg 
cm$^{-2}$ s$^{-1}$.
The extrapolation of the SPL, with $A_s$ and $\Gamma_s$ the best fit values
in Table 2, gives 18.6$\times$10$^{-11}$ erg
cm$^{-2}$ s$^{-1}$. While the near coincidence
is probably fortuitous, in view of variability and the uncertainty on
$\Gamma_s$, it suggests a close connection between the UV and the
soft X--ray excess emission. This result is in agreement with the 
finding by Wang et al. (1998) in a heterogeneous sample of objects
(see also Laor et al. 1997),
but it adds value to it for this particular galaxy, because it is
obtained extrapolating the soft excess, rather than the simple
power law fitted to the ROSAT data. For an approach
similar to ours, based on simultaneous ROSAT and Ginga observations
of other objects, see Walter et al. (1994).  Since Mk 509 for its
luminosity could be ranked among quasars, it is appropriate
to note that $\Gamma_s$ is close to the mean value of 2.8 in the range 
350--1050 A 
found by Zheng et al. (1997) in the composite HST spectrum of
41 radio-quiet quasars.

Another important property of the continuum, seen for the first time
in this object, is the evidence of a cut--off occurring at
$E_f \sim$70 keV. The few other objects in which this property has
been determined have best fit
values greater than 100 keV (Matt 2000, and references therein), the only
exception being NGC 4151 (Jourdain et al. 1992, Zdziarsky et al. 1996,
Piro et al. 1998). Since Mrk 509 shares with NGC 4151 a slope
much harder than the average, although not quite as hard,
this results adds some confidence to the hint of a correlation
between $E_f$ and $\Gamma$ found by Piro (1999).

To assess the issue of the warm absorber, we retrieved
from 
%the Tartarus Database 
{\sc http://tartarus.gsfc.nasa.gov/}
the 0.6-10 keV data of the ASCA observation
(1994, April 29)
investigated by Reynolds (1997) and by George et al. (1998).
The average intensity in the two observations is
the same within 10\%. As a simple compatibility test, we estimated 
the $\chi^2$ when the BMS+SPL+WA model is adopted, with
all parameters equal to the best fit values in Table 2,
except for the normalizations $A_s$, $A$ and $I_k$,
which are left free,
and with the O optical depths first set to zero. We obtain
an unacceptable $\chi^2$=2020/1881. The residuals 
undoubtedly show the signature of the O {\sc vii} edge, and
upon letting the O edges also free, we obtain indeed a pretty acceptable
$\chi^2$=1933/1879, with $\tau$(O {\sc  vii})= 0.20$\pm$0.04 (90\% for
one i. p., $\Delta\chi^2$=2.71), $\tau$(O {\sc  viii})
consistent with zero, a ratio $A_s/A$=1.1, 
within 15\% of the BeppoSAX fit, and an iron line EW $\sim$
110 eV, similar to the BeppoSAX fit.

>From this exercise we would conclude that the two independent
data sets are consistent with a model which includes both 
a ``soft excess", in the form
of a steepening of the power law below about 1 keV, as well as
a warm absorber. Due to the much higher statistics 
and better resolution at the appropriate energies, it is not
surprising that the ASCA data put a stronger and more reliable
constraint on the warm absorber than our data. The
$\tau$(O {\sc  vii})
given above is larger than 0.11$^{+0.03}_{-0.04}$,
the value obtained by Reynolds (1997) jointly with
$N_H=2.1\times10^{20}$ cm$^{-2}$. 
In our exercise an almost identical value, $\tau$(O {\sc  vii})
= 0.10,
is obtained instead when we leave $N_H$ free:
the $\chi^2$ drops to 1850/1878, and $N_H$=3.6$\times10^{20}$ cm$^{-2}$,
but this column is by at least one order of magnitude
incompatible with our data
(see Table 2).
This discrepancy, already noted in the BMS+WA fit (Sect. 3.1),
is probably connected with the 
epoch-dependence in the response of the ASCA-SIS instruments
below 1 keV, which is not yet well understood 
(T. Yaqoob, private communication).

For a slab seen face-on ($i$=0$^{\circ}$), the iron line EW$\sim$100 eV is 
roughly consintent with both a normal abundance and the
solid angle $\Omega\sim\pi$ corresponding to the best fit value of $r$
(Matt et al. 1991, Matt et al. 1997). However,
lacking a constraint on the inclination angle,
and taking into account the uncertainties, the issue
whether $\Omega$ is indeed much less than 2$\pi$,
as it was rather convincingly found to be the case
for the reprocessing in the accretion disk of IC 4329A
(Perola et al. 1999), remains open.

\subsection{MCG 8-11-11}

Our estimate of $r$ for a face--on reprocessing slab is fully
consistent with an $\Omega=2\pi$ geometry. The corresponding line
EW, for iron with solar abundance, should be about 190 eV
(Matt et al. 1997), larger than the best fit value of 130 eV,
but marginally within the errors. The line width is too poorly
constrained for us to elaborate on the contribution by the accretion
disk, and note that also in the ASCA observation (Grandi et al. 1998)
the line is barely resolved. We cannot exclude that a sizeable 
fraction of the reprocessing gas might be located further out. 
In this respect, comparison with the results obtained from
the ASCA+OSSE observation by Grandi et al. (1998) is suggestive,
since at that epoch the source was a factor 2.4 fainter. They
find $\Gamma$ = 1.73$\pm$0.06, $r$ = 1.64$^{+0.88}_{-0.78}$,
 EW = 230$^{+222}_{-99}$ eV
(errors for $\Delta\chi^2$=2.71). None of these estimates differ from 
ours more than the combined errors. Yet the smaller value of $\Gamma$
agrees with an earlier finding (Treves et al. 1990) of a correlation
between $\Gamma$ and flux, while the larger values of both $r$ and EW
could be due to a lag in the response from a reprocessing material
located far away from the variable PL source. More sensitive and better 
resolution observations might confirm this suggestion, which
indicates an $\Omega<<2\pi$ geometry for
the reprocessing in the accretion disk also in this object.

Concerning the absorption, we find for the first time
an admittedly modest evidence of a warm absorber; at the same time
we do not confirm the claim by Grandi et al. (1998) of a column
significantly in excess of $N_{H,g}$, a discrepancy comparable to
that found in Mrk 509 and likely due to
the same cause.

We confirm the presence of the cut--off in the PL, with $E_f$ about
170 keV, smaller than the value 266$^{+90}_{-68}$ keV found by 
Grandi et al. (1998),
but consistent within the combined errors. 

\section{Conclusions}

A particularly interesting result of the present study is the
overall shape of the "direct" continuum in Mrk 509, which
can be empirically described as a combination of two
power laws. The soft power law we detect is, in the first place,
a confirmation of earlier evidence of a low energy excess
(see Sect. 1),
and it seems to us a better description of the excess than
the simple black body adopted by Pounds et al. (1994).
With a slope $\Gamma_s$ about 2.5, it appears as a continuation
into the X-rays of the UV emission, and according to models
sometimes invoked for this component (e. g. by Zheng et al. 1997,
but see also Walter et al. 1994, Fiore et al. 1995, Laor et al.
1997 for a broader discussion of the problem),
it could arise from comptonization in the innermost parts
of the disk in regimes of high accretion rate (e. g. Czerny
\& Elvis 1987, Maraschi \& Molendi 1990). The hard power law
is affected by a high energy cut--off, which can be
empirically described as an exponential factor with $E_f$ about 70 keV.
The currently most popular models for this component involve
comptonization by a hot corona (e. g. Haardt \& Maraschi 1993,
Zdziarsky et al. 1994, Ghisellini et al. 1998), in which the
cut--off is related to the temperature of the corona. 
Our results on Mrk 509 are similar to those on NGC 4151, possibly 
indicating that Seyferts with unusually flat continua feature unusually low 
cut--off energies (Piro 1999). However it is fair to say that proper, 
selfconsistent models need to be used in
the fits in order to recover correctly the physical parameters.
As extensively discussed by Petrucci et al. (2000) in the context of 
a BeppoSAX observation of NGC 5548, the 
exponential cut--off powerlaw is indeed a quite rough approximation to 
realistic componization models. 

Overall the spectrum of Mrk 509 requires a complex structure
for the accretion disk, to explain both the soft and the hard components.
This will be dealt with in a future paper, where we would like to address
the problem in a comprehensive manner for all the objects in the BeppoSAX 
sample of bright Seyfert 1. There we shall include also a more
advanced analysis of the comptonization model, underlying the
continuum shape and in particular the high energy cut--off
seen in MCG~+8-11-11, already discussed in Grandi et al. (1998).

This approach will also attempt to deal selfconsistently with the
"reprocessed" spectral components, namely the amount of continuum
reflection along with the energy/width/intensity of the iron line,
contributed by the matter in the accretion disk.
Here we limit ourselves to note that, although not conclusively
yet, directly from this observation in Mrk 509 and from
comparison with an ASCA observation in MCG~+8-11-11,
there are indications of the accretion disk contributing
to the reprocessing much less than 
one would expect if the solid angle $\Omega$ were 2$\pi$.
Such a situation was also, and rather more convincingly 
found to hold
in IC 4329A (Perola et al. 1999). 

Finally, in both Mrk 509 and MCG~+8-11-11 our observations
present some evidence of a warm absorber. This evidence is
roughly consistent with the more significant one 
found with ASCA in Mrk 509 by Reynolds (1997) and by
George et al. (1998), who did not however confirm in their
analysis the presence of a soft excess. We have shown that the
ASCA data are compatible with the coexistence of the
warm absorber together with the soft excess observed
by BeppoSAX. While doing that, we have noted that the 
interpretation of the ASCA data is complicated by 
a not yet perfectly understood, epoch-dependent,
problem with the response in the ASCA-SIS
low energy channels (T. Yaqoob, private communication): 
the implication (which is
noted to be present, but with less of a consequence, in the ASCA
observation of MCG~+8-11-11, Grandi et al. 1998) is equivalent
to an additional $N_H$ of a few $10^{20}$ cm$^{-2}$, which is
largely incompatible with our data.

\bigskip
\bigskip

%\section*{Acknowledgements}
\small
{\it Acknowledgements.}
The BeppoSAX satellite is a joint Italian--Dutch program. We wish to
thank the Scientific Data Centre for assistance. 
This research has made use of the TARTARUS database, which is 
supported by Jane Turner
and Kirpal Nandra under NASA grants NAG5-7385 and NAG5-7067,
and of the NASA/IPAC Extragalactic Database (NED)
which is operated by the Jet Propulsion Laboratory, California Institute of
Technology, under contract with the National Aeronautics and Space
Administration.
This work
was supported by the Italian Space Agency, and by the Ministry
for University and Research (MURST) under grant COFIN98--02--32.

\end{document}